# Surface-barrier detector with smoothly tunable thickness of depleted layer for study of ionization loss and dechanneling length of negatively charged particles channeling in a crystal

A.V. Shchagin[1,2,*], G. Kube[1], S.A. Strokov[1], W. Lauth[3]

[1]Deutsches Elektronen-Synchrotron DESY, Notkestrasse 85, 22607 Hamburg, Germany
[2]Kharkov Institute of Physics and Technology, Academicheskaya 1, Kharkiv 61108, Ukraine
[3]Institute of Nuclear Physics, Johannes Gutenberg University, J.J.-Becher-Weg 45, 55128 Mainz, Germany

*Corresponding author, e-mail: alexander.shchagin@desy.de

**Abstract**

A new method for the experimental study of ionization loss of relativistic negatively charged particles moving in a crystal in the channeling regime using a semiconductor surface-barrier detector with smoothly tunable thickness of the depleted layer is proposed. The ionization loss can only be measured in the depleted layer of the detector. The thickness of the depleted layer in a flat semiconductor detector can be smoothly regulated by the value of the bias voltage of the detector. Therefore, the energy distribution of the ionization loss of relativistic particles which cross the detector and move in the channeling regime in the detector crystal can be measured along the path of the particles at variation of the bias voltage of the detector. Ionization loss spectra should be different for channeling and nonchanneling particles, and both fractions can be determined. The application of a Si surface-barrier detector-target is considered. Measurements with such a detector would make it possible to obtain the differential by particle path information about the particle's dynamics inside the crystal. The space resolution of this differential measurements can be less than the dechanneling length. Comparison of experimental data with calculations can help to develop a theoretical description of the dynamics of motion of negatively charged particles channeling in a crystal. A better understanding of the dechanneling length properties can be useful in the production of positrons and other particles such as neutrons by an electron beam in crystals, and in the development of crystalline undulators, and at a crystal-based extraction of electron beams from a synchrotron.

## 1. Introduction

After studying the channeling phenomena of low-energy particles reviewed in [1], the research of the channeling of relativistic charged particles began [2 - 4]. It was found that the features of channeling of positively and negatively charged particles are rather different.

Positively charged particles in the channeling regime move between crystallographic planes or rows, at some distance from the atoms of the crystal. Therefore, the probability of the scattering of the particles on crystal atomic nuclei is reduced for channeling particles and they produce less ionization of crystal atoms and have reduced ionization loss. It turned out that ionization loss of relativistic channeling protons [4], as well as positrons [5], is almost two times lower compared to nonchanneling particles. This remarkable property was used for selection of channeling protons in the first [6 - 8] and following [9] experiments on the steering of proton beams by bent crystals. The dechanneling length of ultrarelativistic protons in a Si crystal can be on the order of a few centimeters in the (111) and (110) Si crystallographic planes [9]. At present, channeling of protons in bent crystals is used to extract a part of the proton beam in different accelerator facilities [9].

The situation is different with ionization loss of negatively charged channeling particles. Negatively charged channeling particles move around crystallographic planes or rows in the vicinity of crystal atoms. Therefore, negatively charged channeling particles should experience increased scattering on nuclei of the crystal and produce increased ionization loss. However, no increase in ionization loss at channeling of 1.35 GeV/c $\pi^-$ mesons along the <110> axis or in the main crystallographic planes in Ge detector-target with thickness of 0.67 mm was observed [4]. Besides, no increase of ionization loss of 1.2 GeV electrons channeling along the <111> axis and (110) plane of a 1.6 mm thick Si crystalline detector-target was also observed in [10]. The reason of such unexpected results (in comparison with positively charged channeling particles) could be that the dechanneling length of $\pi^-$ mesons and electrons was much less than the thickness of the detector-targets 0.67 – 1.6 mm in [4, 10]. In this case, the incident channeling $\pi^-$ mesons or electrons quickly leave the channeling regime and pass the most of their path in the detector-target in the nonchanneling regime, and produce ionization loss mainly as nonchanneling particles. But, in the experimental studies at higher energies [11], some increase in the ionization loss was reliably observed at the channeling of 15 GeV/c $\pi^-$ mesons along the <110> crystallographic axes of Si and Ge detector-targets with a thickness of 0.9 mm and 0.74, 2.0 mm respectively (see Figs. 22 – 24 in [11]). The dechanneling length for negatively charged particles was not measured in [4, 10, 11].

Note that the experimental measurements of the ionization loss spectra in [4 - 11] were performed in totally depleted Ge or Si detectors which served simultaneously as crystalline targets and ionization loss detectors. The thickness of the depleted layer in a totally depleted detector, in which ionization losses are measured, is fixed and almost equal to the thickness of the crystalline plate of the detector.

Other methods were also used to measure the dechanneling length of negatively charged particles. In [12], the emission of non-relativistic electrons from the crystal surface at passing of 1.2 GeV electrons through Si crystal was used, where the dechanneling lengths were determined as 39 µm along the <111> axis and 29 µm in the (110) plane. The dechanneling length of particles which emitted channeling or bremsstrahlung radiation was measured in Mainz [13, 14]. Some results of previous measurements and calculations of the dechanneling lengths of electrons with energies up to 1.2 GeV at the crystallographic plane (110) of a flat silicon crystal are presented in Fig. 7 of [15]. One can see a significant disagreement between the theoretical and experimental results. Experimental data give almost constant values of the dechanneling length of about 30 - 35 µm instead of the theoretically predicted linear growth [15, 16] of the dechanneling length as a function of the incident electron energy.

At higher particle energies, the method of observing the angular distribution of 150 GeV/c $\pi^-$ mesons passed through a bent 1.91 mm thick Si crystal along the (110) plane was used in [17], where the dechanneling length of 0.93 mm was determined. A similar method was used in [18], where channeling, volume reflection, and volume capture were studied in the interaction of 3.35 - 14.0 GeV electrons with a 60 µm thick bent Si crystal. The authors obtained almost constant values of the dechanneling length of about 50 – 60 µm for channeling at the (111) crystallographic plane in the whole range of electron beam energies (see experimental data presented in Fig. 19 of [18]).

To our knowledge, the reason of the disagreement between measurements by different methods and some theoretical calculations of the dechanneling length is still not understood. Results of simulations can be found in Ref. [19] and references cited therein. A discussion about the behavior of the dechanneling length can be found, e.g., in [20].

The authors of [21, 22] proposed to measure experimentally the distribution of ionization loss in a crystal to determine the dechanneling length of electrons. They suggested to measure the energy distribution (spectrum) of the ionization loss in crystalline detectors of several fixed thicknesses comparable to the dechanneling length and to compare the measured spectra of ionization loss with the calculated ones. Similar measurements were performed in [4, 10, 11] but for only one thickness of the detector-target which probably sufficiently exceeded the expected dechanneling length of electrons. In particular, the authors of [21] proposed to use spectra of ionization loss of channeling and nonchanneling electrons in only one detector-target of fixed thickness for determining the dechanneling length.

Here we propose to measure the spectra of ionization loss of negatively charged channeling particles in a semiconductor crystalline detector-target with smoothly tunable thickness of the depleted layer and to study the evolution of the distribution of ionization loss as a function of path of particles in the depleted layer. The ionization loss is measured only in the depleted layer of the detector. The thickness of the depleted layer can be controlled by a simple variation of the value of high voltage power supply of the detector. Such research would allow to check the validity of theoretical predictions [15, 16, 21] and to clear up the role of the rechanneling effect. The concept of application of the detector with smoothly tunable thickness of the depleted layer for studying the ionization loss of nonchanneling particles was proposed in [23]. The concept was verified experimentally with 0.975 MeV electrons from a radioactive source [23] and with a 50 GeV proton beam [24]. Such detectors-targets can be used to study the distributions of the ionization loss along the path of positively or negatively charged particles moving in the channeling or nonchanneling regime in a crystal.

## 2. Surface-barrier detector with smoothly tunable thickness of depleted layer

The flat surface-barrier Si detector is shown in Fig. 1. It consists of a silicon single crystal plate. The front surface of the detector is usually covered with Au layer of the thickness about hundreds of Angstroms. The back surface is usually coated with an Al layer. The Au and Al layers are the contacts of the detector. When the bias voltage is applied to the detector, the depleted layer appears near the gold contact. The thickness $d$ of the depleted layer depends on the applied bias voltage [25, 26]

$$d = \sqrt{\frac{2\varepsilon\varepsilon_0}{eN}(U+U_K)}, \qquad (1)$$

where $\varepsilon = 11.9$ is the dielectric permittivity of the Si crystal, $\varepsilon_0 = 8.85 \cdot 10^{-12} \frac{F}{m}$ the dielectric permittivity of free space, $-e$ the electron charge, $N$ the density of acceptors or donors, $U$ the bias voltage, and $U_K \approx 0.5$ V for a silicon detector [26]. The bias voltage is applied to the depleted layer and an electric field appears in the depleted layer. The thickness of the depleted layer can be regulated by the value of the bias voltage according to formula (1) and checked via measurements of the capacitance of the detector, as it was done and described in details in [23].

The depleted layers of thickness $d_1$ and $d_2$ is shown in gray color in Fig. 1a and Fig. 1b respectively. The thickness of the depleted layer is $d_2 > d_1$ because $|U_2| > |U_1|$, see Eq. (1). Also, the thickness of the layer equal to the dechanneling length of the particles $L_d$ is shown in Fig. 1. The dechanneling length $L_d$ is the same in both panels in Fig. 1. The thickness of the depleted layer is $d_1 < L_d$ in Fig. 1a, but the thickness of the depleted layer is $d_2 > L_d$ in Fig. 1b.

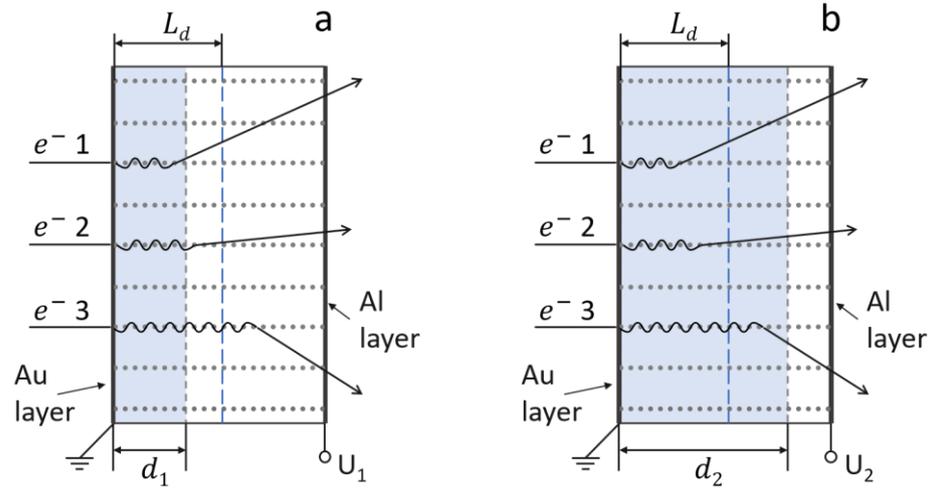

*Fig. 1. The flat surface-barrier semiconductor detector-target with channeling of negatively charged particles. The depleted layer is shown in gray color. The regulated thickness of the depleted layer d may be smaller ($d_1$ in panel a) or larger ($d_2$ in panel b) than the dechanneling length $L_d$ depending on the value of the bias voltage U. The incident negatively charged particles are channeling at the crystallographic planes or along the rows which are shown as dotted lines. The electrons #1, 2 are dechanneling at a distance less than the dechanneling length $L_d$, the electron #3 is dechanneling at a distance exceeding the dechanneling length $L_d$ in both panels. In panel a, the electron #1 is dechanneling within the depleted layer of thickness $d_1$, electrons #2, 3 are dechanneling outside the depleted layer. In panel b, all electrons are dechanneling within the depleted layer of thickness $d_2$.*

When a high-energy charged particle crosses the detector, it produces electron-hole pairs along its entire path in the silicon crystal. The number and charge of electron-hole pairs are proportional to the ionization loss of the particle in the material of the detector. The charge of the part of the electron-hole pairs generated only in the depleted layer is collected by the electric field in the depleted layer. This current pulse is integrated in a charge-sensitive preamplifier and fed to the data acquisition system. The charge of the electron-hole pairs which are produced in the non-depleted layer of the detector is not collected and measured. The registered spectrum of ionization loss contains a single asymmetric so-called Landau peak [27] of ionization loss of particles in the depleted layer, if the particles move in the non-channeling regime. The energy resolution with a surface-barrier detector can be a few keV [23, 24], that is much less than the width of the Landau peak. The energy resolution can be improved by using a detector-target with minimal sensitive area and cooling the detector preamplifier assembly to a temperature of about -100°C. The measured spectra of ionization loss of nonchanneling particles in the depleted layers of different thicknesses can be found, e.g., in [23] and in Fig. 3 in [28] (electrons) and in [24, 29] (protons). But the shape of the ionization loss spectrum of channeling particles can be more complex and contain more than one spectral peak [5, 6, 21].

In real experiments [23, 24], the thickness of the depleted layer varied approximately by a factor of 2 - 2.8 due to variation of the bias voltage by a factor of approximately 4 - 6. The possibility to measure ionization loss spectra in a thinner depleted layer is restricted. The restriction is due to two reasons. The first reason is that the detector capacitance increases with the reduction of the depleted layer thickness which leads to an increase of the noise in spectra. The second reason is that the energy loss, and thus the signal height, becomes smaller as the thickness of the depleted layer decreases. As a result of both reasons, the Landau peak in the spectra penetrates into the noise area and its observation is difficult due to interference with the noise. To overcome the restriction, one can use the detector of minimal area to reduce its capacity. The optimal transverse size of the detector-target should be slightly larger than the transverse size of the incident particle beam to avoid edge effects from the particles which can cross the edges of the detector-target.

The desirable maximum thickness of the depleted layer at the maximum bias voltage can be achieved by choosing the right detector with appropriate density of acceptor or donor atoms in the Si crystal of the detector. Flat Si detectors usually have the maximum depleted layer thicknesses from tens of micrometers to millimeters at maximum bias voltage of up to hundreds of Volts [30]. Such thicknesses of the depleted layer well match to expected values of the dechanneling length from about 30 µm [15] to about 1 mm [17, 21].

The absolute value of the depleted layer thickness can be measured in three ways: i.) by measuring the capacity of the detector, as noted above; ii.) by measuring the Landau peak energy in a beam of nonchanneling particles; iii.) by using the restricted path of alpha particles from a radioactive source in the Si detector (for depleted layer thicknesses of about 30 – 40 micrometers).

## 3. Channeling in silicon detector-target

Commercially available Si surface-barrier detectors are produced from a Si single crystal. Usually, the <111> crystallographic axis and three (110) crystallographic planes are almost perpendicular to the front surface of the detector. The channeling conditions should be the same in the depleted and non-depleted layers because the depletion does not change the atomic and nuclear structure of the crystal. The variation of the depleted layer thickness should not change the channeling conditions either. Therefore, such detectors can be used in experiments on measuring the ionization loss spectra of particles channeling along the Si <111> crystallographic axis and at Si (110) crystallographic planes if the detector is installed almost perpendicularly to the axis of the particle beam, as it is shown in Fig. 1. The dead layer at the front surface of the detector is typically less than 1 µm thick and cannot sufficiently increase the divergence of the incident particle beam.

The critical Lindhard angle $\theta_L$ is

$$\theta_L = \sqrt{\frac{2U_0}{pV}} \qquad (2)$$

where $U_0$ is the depth of the potential well in the crystal in the direction perpendicular to the plane or axis, which is about 20 eV for the Si (110) and Si (111) planes [9], $p$ and $V$ are the momentum and velocity of the incident particles. For example, $\theta_L \approx 0.2$ mrad for electrons or positrons with the energy of 1 GeV for Si (110) and Si (111) planes.

The detector-target should be mounted on a goniometric stage and installed in the particle beam. The angular accuracy of the goniometric stage should be less than $\theta_L$. The incident electron or positron beam should have a divergence less than $\theta_L$ for research the channeling phenomena. The particle beam current of the incident beam should be not more than a few thousands particles per second in continuous mode to avoid distortions of the measured spectra due to pile-up effects in the spectrometer.

The maximum thickness of the depleted layer of the detector-target at maximum bias voltage can be about 40 - 50 µm for an expected dechanneling length of 30 - 35 µm [15] or about 1.5 mm for an expected length of about 1 mm [17, 21]. The thickness of the depleted layer can be reduced at the reduction of the bias voltage. The alignment of the crystal in the beam can be performed by observation scattered electrons with an ionization chamber as it was done in [31]. Also, one can try to perform the alignment via observations of the ionization loss by controlling the ratio of the number of particles in the main Landau peak and in the low-energy tail of the peak in the case of positively charged particle beams, or in the high-energy tail in the case of negatively charged particle beams such as electrons. As an alternative method of the alignment, parametric X-ray radiation (PXR) can be used [32]. In this case, PXR from the crystallographic planes of the detector-target should be observed in the backward hemisphere by an X-ray spectrometer.

The detector-target should be aligned with use of the goniometric stage to provide the channeling regime in the crystal for particles of the incident beam. After that, one can measure spectra of ionization loss as a function of the depleted layer thickness and analyze them.

**4. Discussion**

Let us discuss peculiarities of the proposed method.

1. It seems that the ionization loss is rather convenient for the study of the dechanneling because the most probable energy loss (energy of the Landau peak) is practically independent for electron energies above 1 MeV due to the Fermi density effect, at least for nonchanneling particles (see, e.g., Fig. 6 in [23]). Therefore, the method can be applied at any incident electron beam energy exceeding 1 MeV. A similar situation is valid for other relativistic charged particles.

2. Ionization loss spectra in the channeling regime should contain two spectral peaks. The first peak is the same Landau peak which is produced by nonchanneling particles. The second peak with reduced energy can be associated with the channeling positively charged particles, as it was observed for positrons [5] and protons [6 - 9]. In the case of an incident beam of negatively charged particles, one can expect the appearance of a second smoothed peak with increased energy relative to the Landau peak, as it was theoretically predicted in [21] and shown in Fig. 2 [21]. Thus, the contributions of channeling and nonchanneling fractions of the incident beam of electrons or positrons can be distinguished.

3. The results of measurements of ionization loss spectra should be independent on the emission or non-emission of channeling radiation by some parts of the relativistic channeling particles. This is because the emission of channeling radiation leads to a reduction of the particle energy only, but the particle remains in the channeling regime. The ionization loss should not be changed due to the Fermi density effect (see item 1).

4. The ionization loss should be the same for particles channeling from the beginning of their motion in the crystal or rechanneling inside the crystal. This peculiarity can help to clear up the role of rechanneling of particles in the crystal.

5. The measurement of the ionization loss in different thicknesses of the depleted layer can be performed in only one immovable detector-target with preliminary fixed alignment.

6. The spectrum of ionization loss measured at only one fixed bias voltage and depleted layer thickness can give information about channeling and nonchanneling fractions of the beam averaged over the entire path of the particles in the depleted layer. But two spectra measured with two different thicknesses of the depleted layer can provide information about processes in the layer inside of the crystal which is between two edges of the depleted layers. This is a unique possibility to extract information about dechanneling or rechanneling processes in a layer inside the crystal. For example, two spectra of ionization loss measured at bias voltages $U_2$ and $U_1$ can give the information about processes in the layer between edges of depleted layers $d_2$ and $d_1$ of thickness $d_2 - d_1$, see Fig. 1. In other words, it is possible to extract differential by particle path data which can give access to the particle dynamics inside the crystal. The space resolution of this differential measurements can be less than the dechanneling length at $d_2 - d_1 < L_d$, especially at high particle energies.

7. A semiconductor detector-target based on different crystalline materials, for instance germanium, can be used in research if the thickness of the depleted layer can be regulated by the bias voltage in such detector.

8. Probably, a bent silicon crystal similar to the one used, e.g., in [18] can serve as a surface-barrier detector for studying the dechanneling length of negatively charged particles and other related phenomena. For instance, ionization loss at volume reflection and volume capture of particles in a bent crystal can be measured and investigated. Note, that similar detectors were installed on the bent Si crystal plates and were successfully used in the first experiments on proton channeling phenomena in [6 - 8].

9. The detector-target can be placed in the beam with its front surface as shown in Fig. 1. In this case one can observe ionization loss of particles starting their motion in the crystal. Also, the detector-target can be placed in the beam with its rear surface. In this case one can observe the ionization loss of particles finalizing their motion in the crystal. Such installation can be useful for research of rechanneling phenomena.

## 5. Conclusion

In this work we propose a new method for further experimental studies of dechanneling phenomena and to obtain differential information about dynamics of the particle in crystal. The peculiarities of the proposed here and in [21] experimental research methods would allow to clarify the situation with the dechanneling length of electrons. A better understanding of the dechanneling length properties can be useful in the production of positrons [33, 34] and other particles such as neutrons by an electron beam in crystals, and in the development of crystalline undulators (see, e.g., [35]), and at a crystal-based extraction of electron beams from a synchrotron [36].

**Acknowledgments**

The authors are thankful to H. Backe, S.V. Trofymenko, and I.V. Kyryllin for useful discussions. A.V. Shchagin is grateful to DESY Hamburg (Germany) and Helmholtz Association (HGF) for granted asylum after fleeing the war in Ukraine, outstanding support and provided funding from the Initiative and Networking Fund under the contract number GI-022.